\documentclass[apjl]{emulateapj}
\newcommand{\na}{New Astro.}

\begin{document}

\title{Effect of Hot Baryons on the Weak-Lensing Shear Power Spectrum}
\author{Hu Zhan and Lloyd Knox} 
\shortauthors{Zhan \& Knox}
\affil{Department of Physics, University of California, Davis, CA 95616}
%\email{zhan@physics.ucdavis.edu}
%\email{lknox@physics.ucdavis.edu}
\email{zhan@physics.ucdavis.edu, \\ 
\phantom{Electronic address: }lknox@physics.ucdavis.edu}

\begin{abstract}
We investigate the impact of the intracluster medium on the
weak-lensing shear power spectrum (PS). Using a halo model we find
that, compared to the dark matter only case, baryonic pressure leads
to a suppression of the shear PS on the order of a few percent or 
more for $l \gtrsim 1000$. Cooling/cooled baryons and the 
intergalactic medium can further alter the shear PS. Therefore, the 
interpretation of future precision weak lensing data at high
multipoles must take into account the effects of baryons.
\end{abstract}


\section{Introduction} \label{sec:intr}
Weak gravitational lensing is a promising technique for precision
cosmology.  Its chief advantage over other techniques comes from its
sensitivity to the total density field, rather than just the baryonic
component.  Since the total density is mostly due to the dark matter,
which only interacts gravitationally, the statistical properties of
the total density field can be calculated \emph{ab initio} with much 
higher accuracy than can observables that are more directly dependent 
on baryons.  Combined with the enormous statistical power of proposed
survey projects, such as the Large Synoptic Survey Telescope\footnote{
\url{see http://www.lsst.org.}} and 
\emph{SuperNova/Acceleration Probe}\footnote{\url{see http://snap.lbl.gov.}}, 
weak lensing is thought to be capable of determining (some) cosmological 
parameters to a few percent level \citep[e.g.][]{rmr04}.

Many studies of weak-lensing statistics assume baryons to trace dark 
matter exactly \citep[e.g.][]{bre00, chm00, hw01, sk03}. This allows 
one to predict the weak-lensing shear power spectrum (PS) from the 
nonlinear dark matter PS that is well calibrated using 
$N$-body simulations \citep{pd96, mf00, spj03}.

However, more than 10\% of the total mass is in baryons, which do not
follow dark matter exactly on small scales.
% \citep{gh98, z04}.
Therefore the total mass distribution and its PS will deviate somewhat
from those in a universe that has the same initial conditions except
that all the mass is in dark matter.\footnote{One may think of this as
an $N$-body simulation of dark matter that has the same parameters and
initial mass PS (including baryonic features such as due to
acoustic oscillations) 
as a full hydrodynamical simulation.} Such a deviation in the mass PS will
result in a difference in weak lensing statistics.  Eventually, it
will be necessary to account for the baryonic influence on the shear
PS as we reach for the projected statistical power of weak lensing
surveys.

Baryons exist mainly in three categories: (1) cooling/cooled baryons
such as in stars and galaxies, (2) hot baryons or the intracluster
medium (ICM), and (3) the intergalactic medium (IGM). 
The cooling/cooled baryons alter the statistics of gravitational 
lensing \citep{w04}, because they can condense to a much 
denser state than dark matter, which in turn modifies the dark matter 
distribution \citep{kkz04}.  The ICM
is smoother than cooling/cooled baryons, so that the effect might be
much less pronounced \citep{w04}. However, the ICM greatly out-weighs 
the baryons in the galaxies. Therefore, it could still have a 
significant impact on the shear PS. Indeed, we find at 
$l \la 3000$ the ICM has a larger
impact on the shear PS than do the cooling/cooled baryons.

For completeness we note here that the IGM is much more diffuse
than the other two,
but it is the largest reservoir of baryons. Since the density
fluctuations on the angular scales we consider are dominated by
high-density and massive objects, we neglect the IGM for the present,
but plan to study its effect using hydrodynamical simulations in
the future.

In this {\it Letter} we demonstrate the effect of hot baryons on the
weak-lensing shear PS. Assuming an isothermal $\beta$-model
\citep{cf76}, we are able to extend the
halo model \citep[for a review, see][]{cs02} to include hot baryons in
the total mass PS, which is then used to calculate the shear PS. Aside 
from the uncertainties of shear statistics due to halo parameters
\citep{tj03}, this method is not sufficiently accurate for interpreting 
future data, because we have not included all the
baryons, and because we have made simple associations of the ICM profile 
with its dark matter halo. Nonetheless, it provides a framework for 
analytically estimating the effect of uncertainty in the clustering 
properties of hot baryons.

\section{Mass Power Spectrum} \label{sec:mps}
Realistic halo profiles \citep[e.g.][]{nfw96, mqg99} enable the halo 
model to reproduce the nonlinear dark matter PS.
With the one-halo PS $P^{\rm 1h}_{\rm dm}(k)$ and 
two-halo PS $P^{\rm 2h}_{\rm dm}(k)$, the dark matter PS $P_{\rm dm}(k)$ 
can be approximated by \citep{ps00, mf00, s00, ssh01}
\begin{eqnarray} \label{eq:haloPS}
P_{\rm dm}(k) &=& P^{\rm 1h}_{\rm dm}(k) + P^{\rm 2h}_{\rm dm}(k), \\
P^{\rm 1h}_{\rm dm}(k) &=& \int f(\nu){\rm d}\nu \,
%\frac{m_{\rm d}(\nu)}{\bar{\rho}_{\rm d}} 
m_{\rm d}(\nu) \left | u_{\rm d}[k,m_{\rm d}(\nu)] \right |^2
/\bar{\rho}_{\rm d}, \nonumber \\
P^{\rm 2h}_{\rm dm}(k) &=& P_{\rm L}(k) \Big\{ \int f(\nu)b(\nu)
{\rm d}\nu \, u_{\rm d}[k, m_{\rm d}(\nu)]\Big\}^2, \nonumber
\end{eqnarray}
where $\bar{\rho}_{\rm d}$ is the mean density of the universe at the 
present, $u_{\rm d}[k,m_{\rm d}(\nu)]$ is the normalized profile of a 
dark matter halo with mass $m_{\rm d}$ in Fourier space 
\citep[e.g.][]{cs02}, and $P_{\rm L}$ is the linear 
mass PS. The dark-matter density profile is cut off at the
virial radius. The height of the density peak, $\nu$, is defined as
\begin{equation} \label{eq:nu}
\nu = [\delta_{\rm c}(z) / \sigma(m_{\rm d})]^2,
\end{equation}
where $\delta_{\rm c}(z) \simeq 1.686\Omega^{0.0055}_{\rm m}(z)$ 
\citep{h03}  is the overdensity of a spherical region that collapses at 
redshift $z$, $\Omega_{\rm m}(z)$ is the ratio of matter density to 
critical density at $z$, and $\sigma(m_{\rm d})$ is the rms value of the 
density contrast $\delta \rho_{\rm d}/\rho_{\rm d}$ within a radius of 
$(3m_{\rm d}/4\pi\bar{\rho}_{\rm d})^{1/3}$ at $z$. Equation 
(\ref{eq:nu}) also defines $m_{\rm d}(\nu)$. The functions 
$f(\nu)$ and $b(\nu)$, related to the halo mass function and bias, 
respectively, have the forms \citep{st99}
\begin{eqnarray} \nonumber
\nu f(\nu) &=& A (1 + \nu_1^{-p})\,\nu_1^{1/2}e^{-\nu_1/2}, \\ \nonumber
b(\nu) &=& 1 + (\nu_1-1)\,\delta_{\rm c}^{-1} +  2p\, 
\delta_{\rm c}^{-1} (1+\nu_1^p)^{-1},
\end{eqnarray}
with $\nu_1 = 0.707 \nu$ and $p = 0.3$. The normalization constant $A$ 
satisfies the constraint $\int f(\nu) {\rm d} \nu = 1$.

Equation (\ref{eq:haloPS}) can be extended to include hot baryons:
\begin{eqnarray} \label{eq:bps}
%P(k) &\simeq& f_{\rm d}^2 P_{\rm dd}(k) + 
%  2 f_{\rm b} f_{\rm d} P_{\rm bd}(k) + f_{\rm b}^2 P_{\rm bb}(k), \\
P(k) & = & \sum_{ij} f_i f_j P_{ij}(k), \\
P_{ij} &=& P_{ij}^{\rm 1h} + P_{ij}^{\rm 2h}, \nonumber \\ \nonumber
P_{ij}^{\rm 1h} & = & \int f(\nu) {\rm d} \nu \,m_{\rm d}(\nu)
%\frac{m_{\rm d}(\nu)}{\bar{\rho}_{\rm d}} 
u_{i}[k,m_{\rm d}(\nu)] 
u_j[k, m_{\rm d}(\nu)] / \bar{\rho}_{\rm d}, \\
P_{ij}^{\rm 2h} & = & P_{\rm L}(k) \int f(\nu) b(\nu) {\rm d} \nu 
u_{i}[k,m_{\rm d}(\nu)] \nonumber \\ & & \quad \times 
\int f(\nu') b(\nu') {\rm d} \nu' u_j[k, m_{\rm d}(\nu')], \nonumber
\end{eqnarray}
where the subscripts $i,j = {\rm d}$ for dark matter and ${\rm b}$ for 
baryons. The coefficients, $f_{\rm d}$ and $f_{\rm b}$, are the 
fractions of dark matter and baryons in total halo mass, respectively. 
Obviously, $P_{ij}(k) = P_{ji}(k)$, $f_{\rm d} + f_{\rm b} = 1$, and 
equation (\ref{eq:haloPS}) is a special case of equation (\ref{eq:bps})
when $f_{\rm b} = 0$. 

For equation (\ref{eq:bps}) we have assumed a fixed gas-mass fraction 
$f_{\rm b}$. In principle, there is a distribution of $f_{\rm b}$ for 
a given halo mass, and this distribution can be dependent on halo mass 
\citep[or on the ICM temperature, see][hereafter MME99]{ae99, mme99}. 
Although the dependence is found to be weak for clusters 
\citep[][hereafter OM04]{om04}, it may not be true for low mass systems. 
Hence, an average over the distribution should be incorporated 
to calculate the PS. We assume for  
convenience that $f_{\rm b}$ equals the cosmic mean baryon fraction 
of $0.13$. 

The Fourier-space baryon profile $u_{\rm b}(k, m_{\rm d})$ is 
calculated from the $\beta$-model, 
%\[
%\rho_{\rm b}(r) = \rho_{\rm b0}(1+r^2/r_{\rm c}^2)^{-3\beta/2},
%\]
$\rho_{\rm b}(r) = \rho_{\rm b0} (1+r^2/r_{\rm c}^2)^{-3\beta/2}$,
where $\rho_{\rm b0}$ is the central baryon density, $r_{\rm c}$ the 
core radius, and $\beta = 0.4$--$1.2$ 
\citep[MME99; OM04;][]{mdm03, etb04}. 
The baryon profile is also cut off at the virial radius. 
Assuming hydrostatic equilibrium and the NFW halo profile \citep{nfw96}, 
\citet{mss98} find that the best-fitting 
$\beta$-model has a core radius about one-fifth of the scale radius,
$r_{\rm s}$, of the NFW profile. We choose $\beta = 0.7$ and  
$r_{\rm c} = 0.22 r_{\rm s}$ to be the fiducial model and allow the 
parameters to vary. Note that the polytropic gas profile \citep{ks01}
is slightly smoother than our fiducial model and it gives similar 
results.

\begin{figure}
\centering
%\epsscale{0.5}
%\plotone{f1}
\includegraphics[width=75mm]{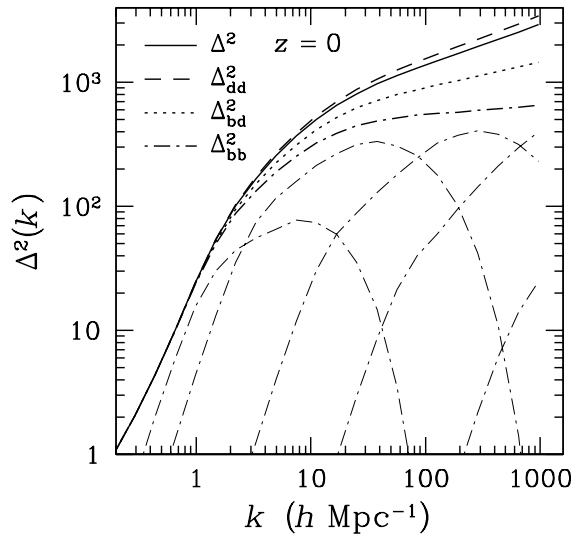}
\caption[f1]{
Halo-model mass PS at $z = 0$. The total PS
$\Delta^2(k)$ (\emph{solid line}) is calculated using equation 
(\ref{eq:bps}) as a weighted sum of the dark matter
PS $\Delta^2_{\rm dd}(k)$ (\emph{dashed line}), the baryon-dark 
matter cross PS $\Delta^2_{\rm bd}(k)$ (\emph{dotted line}), and
the baryon PS $\Delta^2_{\rm bb}(k)$ (\emph{dash-dotted line}). 
From left to right, the thin dash-dotted lines represent one-halo
contributions to the baryon PS from mass ranges of
$10^{17} M_\sun \ge m_{\rm d} > 10^{14} M_\sun$, 
$10^{14} M_\sun \ge m_{\rm d} > 10^{12} M_\sun$, 
$10^{12} M_\sun \ge m_{\rm d} > 10^{10} M_\sun$, 
$10^{10} M_\sun \ge m_{\rm d} > 10^{7} M_\sun$, and
$10^{7} M_\sun \ge m_{\rm d} \ge 10^{4} M_\sun$. 
\label{fig:halops}}
\end{figure}

Fig.~\ref{fig:halops} shows the fiducial model PS at $z = 0$ in 
dimensionless form, i.e. $\Delta^2(k) = k^3 P(k)/2\pi^2$. We have 
assumed a flat universe with $\Omega_{\rm m} = 0.3$, $\Gamma = 0.18$, 
$\sigma_8 = 0.85$, and $n = 1$, where $\Gamma$ is the shape parameter
of the PS, $\sigma_8$ is the rms value of the density contrast within
a radius of 8 \mbox{$h^{-1}$Mpc}, and $n$ is the power spectral index.
Because, for the fiducial model, the baryon profile is smoother than 
the dark matter profile, the baryon PS $\Delta_{\rm bb}^2(k)$ and 
baryon-dark matter cross PS $\Delta_{\rm bd}^2(k)$ become 
significantly lower than the dark matter PS $\Delta_{\rm dd}^2(k)$ on 
small scales ($k > 1$ \mbox{$h$ Mpc$^{-1}$}) where the one-halo term 
dominates. At higher redshift, this departure occurs at a larger 
wavenumber. Consequently, the total mass PS is lower than what it 
would be if the hot baryons are replaced by dark matter.

From the break-down of one-halo contributions to the baryon PS, 
also shown in Fig.~\ref{fig:halops}, one sees that each group of halos 
dominates a certain range of scales. 
%We have included very low mass  ($m_{\rm d} < 10^{10} M_\sun$) halos 
%in our calculations. 
The $\beta$-model might not apply to halos with 
$m_{\rm d} < 10^{10} M_\sun$, but 
since they contribute little to the mass PS at $k < 100$ 
\mbox{$h$ Mpc$^{-1}$} -- the most important scales to shear PS at $l$ 
less than a few thousand -- our conclusions will not be affected.

\section{Shear Power Spectrum} \label{sec:sps}
\begin{figure} 
\centering
%\epsscale{1}
%\plotone{f2}
\includegraphics[width=75mm]{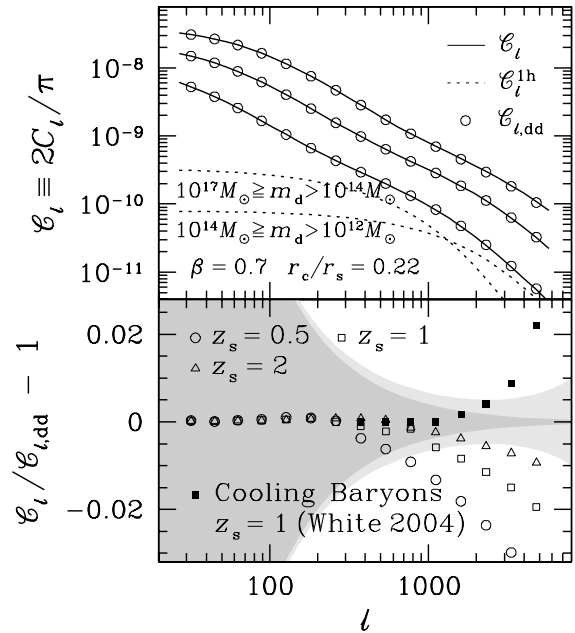}
\caption[f2]{
\emph{Upper panel}: Shear PS with and without 
baryons (\emph{solid lines and circles, respectively}). From bottom to top, 
the solid lines and circles assume sources to be on a thin slice at 
$z = 0.5$, 1, and 2. The dotted lines
represent one-halo contributions to the total shear PS 
$\mathcal{C}_l$ at $z = 0.5$. Contributions from halos with mass less 
than $10^{12}M_\sun$ are below $10^{-12}$.
The fractional difference are shown in the lower panel. The dark 
grey band is the one $\sigma$ statistical error on a band of width $l/4$ 
in the sample variance limit for a survey with $f_{\rm sky}=0.25$.
The light grey band includes shape noise for the $z_s=1$ PS
assuming $\bar n$ = 10/arcmin$^2$ and $\gamma_{\rm rms}=0.2$.

\label{fig:shearps}}
\end{figure}

With the Limber approximation and the assumption that sources are on a 
thin slice at $z$, the shear PS $C_l$ in a flat universe is given by 
\citep{bre00, h00, bs01}
\begin{equation}
C_l = \frac{9}{4} \left(\frac{H_0}{c}\right)^4
\left(\frac{\Omega_{\rm m}}{D_{\rm s}}\right)^2 \int_0^{D_{\rm s}} 
{\rm d} D \left(\frac{D_{\rm s} - D}{a}\right)^2 P(k; z),
\end{equation}
where $H_0$ is the Hubble constant at $z = 0$, $c$ is the speed of 
light, $D$ and $D_{\rm s}$ are the comoving distance of the lens and 
sources, respectively, $a = (1+z)^{-1}$, and $k = l / D$. We define 
$\mathcal{C}_l \equiv 2 C_l / \pi$, which is the contribution to the 
variance of the deflection angle from logarithmic intervals in $l$. The 
results are shown in Fig.~\ref{fig:shearps} for the fiducial model. 
Since the shear PS is mostly affected by the linear and quasi-linear 
part of the mass PS, the effect of hot baryons on the shear PS is less 
than a few percent for $l \lesssim 3000$. As redshift increases, the
same $l$ corresponds to smaller wavenumbers, meanwhile the PS is more 
linear at the same wavenumber. In addition, hot baryons make less of a 
difference in the mass PS at higher redshift. Thus, the effect of the 
ICM on the shear PS reduces with increasing redshift. 

The effect of the ICM on the shear PS is large enough that it will
need to be addressed if the cosmological parameter errors forecasted
in, e.g., \citet[with $l_{\rm max}=1000$]{sk03}, \citet[with $l_{\rm
max} = 3000$]{h02} and \citet[with $l_{\rm max} = 2\times
10^5$]{rmr04} are to be realized.  This is clear from the lower panel
of Fig.~\ref{fig:shearps} where departures outside the shaded region
are larger than the statistical error for a fiducial survey with
parameter values as given in the caption.  The statistical error in
${\cal C}_l$ in a band of width $l/4$ is given by
\begin{equation}
\Delta {\cal C}_l =  0.004 \left(\frac{1000}{l}\right)
\left(\frac{0.25}{f_{\rm sky}}\right)^{1/2} \left({\cal C}_l +
\frac{2}{\pi} \frac{\gamma^2_{\rm rms}}{\bar n}\right),
\end{equation} 
where $f_{\rm sky}$ is the fraction of sky covered, $\bar n$
is the surface number density of sources with measurable shapes,
and $\gamma^2_{\rm rms}$ is the variance of the ``shape noise''.  

From the break-down of the contributions from different mass halos, one 
sees that the two-halo term in the mass PS is the largest contributer 
at large angular scales, and only those halos with
$m_{\rm d} > 10^{12}M_\sun$ are important to the shear PS at $l$ around
a few thousand.

Observationally, the $\beta$ parameter and the core radius assume a range 
of values. In fact, the baryon profile becomes more compact than the dark 
matter profile if $\beta$ is larger and the core-radius-to-scale-radius 
ratio is smaller. This can drive the baryon-dark matter cross PS and 
baryon PS higher than the dark matter PS, and results in an increase in 
the shear PS. On the other hand, the entropy floor \citep{pcn99} can 
increase the core radius of the baryon profile in low mass systems
\citep{tn01, hc01} and reduce the total mass PS and shear PS.

\begin{figure*} 
\centering
%\epsscale{1}
%\plotone{f3}
\includegraphics[width=6in]{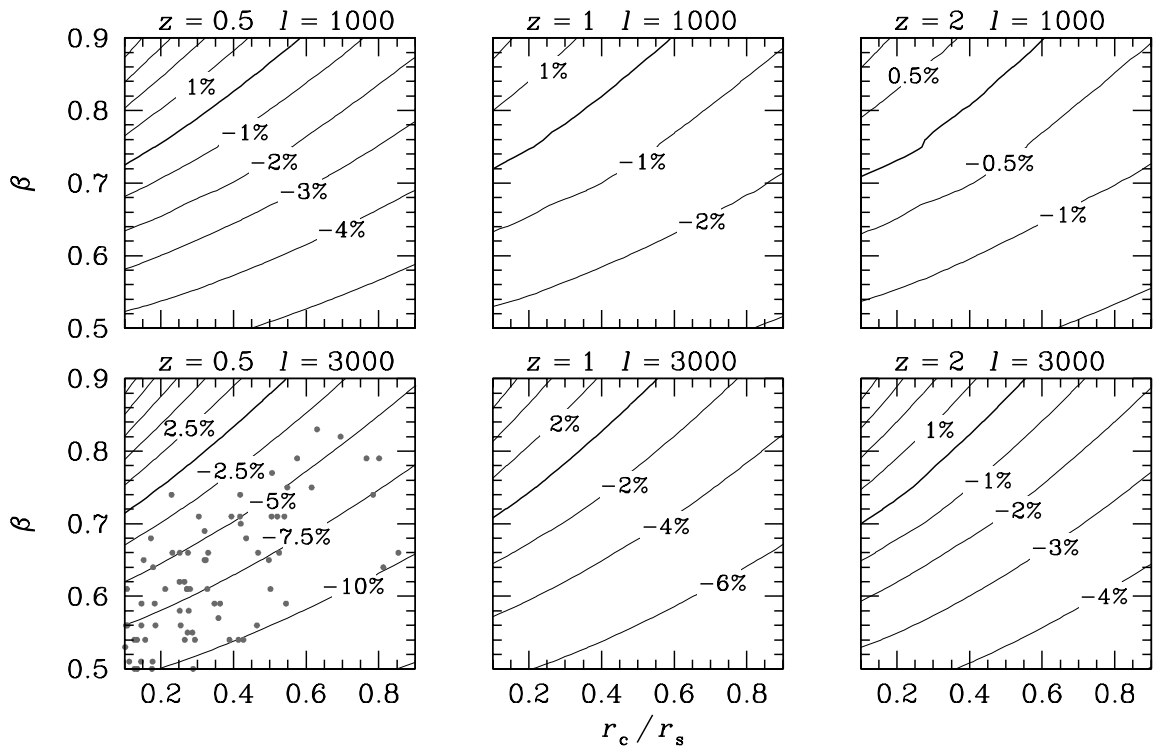}
\caption[f3]{
Contours of the fractional difference, 
$\mathcal{C}_l / \mathcal{C}_{l, \rm dd} - 1$, in 
$\beta$--$r_{\rm c}/r_{\rm s}$ space. Grey dots in the lower left 
panel correspond to the observed values of $\beta$ and 
$r_{\rm c}/r_{\rm s}$ from MME99 and OM04.
\label{fig:shpserr}}
\end{figure*}

Fig.~\ref{fig:shpserr} explores the relative changes in the shear PS in 
the $\beta$--$r_{\rm c}/r_{\rm s}$ space for $l = 1000$ and $3000$. We
see that given the observational uncertainties about the $\beta$-model 
parameters, hot baryons could pose a significant challenge to precision 
cosmology. 

We are optimistic that the challenge can be met, at least for $l \la 3000$,
by a combination of hydrodynamical simulations (which in principle can 
track hot baryons quite accurately), and observations  of hot gas via 
X-ray emission and the Sunyaev-Zel'dovich (SZ) effect.  Fortunately, it is
only very massive halos that are important at $l \la 3000$, 
and these are just the ones easiest to study in X-ray and SZ. Planned
SZ surveys such as South Pole Telescope, Atacama Cosmology Telescope, 
Atacama Pathfinder Experiment and Sunyaev-Zel'dovich Array will have 
sufficient angular 
resolution and observe a sufficiently large number of clusters
to guide the development of a statistical model of ICM profiles.  
Observing strategy can also, to some degree, mitigate the difficulties 
posed by baryons.  Their importance at small scales
argues for large sky coverage, to reduce sample variance
(thereby improving the ${\cal C}_l$ measurement on all scales), 
rather than depth and angular resolution, which reduce the shape noise
that is only important at small scales. 

\section{Discussion and Conclusions} \label{sec:con}
We have extended the halo model to calculate the total mass PS of dark 
matter and hot baryons. Given the observed properties of the ICM, we 
find the total mass PS is considerably lower than the dark matter PS on 
small scales. This leads to a few percent reduction of the shear PS 
at $1000 \lesssim l \lesssim 3000$ compared to what it would be if 
baryons traced dark matter exactly.  The effect grows with 
increasing $l$.  Further, for $l \ga 3000$  an effect 
(with opposite sign, see Fig.~\ref{fig:shearps}) also becomes 
important due to cooling/cooled baryons \citep{w04}.

So far, we have not considered the IGM 
\citep[including the warm-hot IGM,][]{dco01}, 
which contains roughly two thirds of the baryons at $z = 0$. Using
hydrodynamical simulations \citep[TreeSPH,][]{ddh97} that incorporate
cooling, heating, star formation, and feedback, \citet{z04} obtains a 
total baryon PS that is lower than what has been shown in 
Fig.~\ref{fig:halops}.
Therefore, when all the baryons are accounted for, the effect on weak
lensing statistics could be even greater. A detailed investigation
using hydrodynamical simulations will be necessary to assess the total
baryon effect.  In conclusion, for precision cosmology
with weak lensing to live up to its promise, we will have to pay
attention to the modeling of baryons, and not just the dark matter.

\acknowledgements
%We thank A. Stebbins for a useful conversation and J.~A. Tyson for 
%commenting on the manuscript. We also thank C. Fassnacht, L. Lubin, 
%and M. White for discussions on the ICM profile.
%This work was supported by the National Science Foundation under 
%Grant No. 0307961 and NASA under grant No. NAG5-11098

We thank C. Fassnacht, L. Lubin, A. Stebbins, J.~A. Tyson and M. White 
for useful discussions. This work was supported by NSF under 
Grant No. 0307961 and NASA under Grant No. NAG5-11098

\end{document}